\documentclass[12pt]{article}
\usepackage[hmargin=0.7in,vmargin=1.0in]{geometry}
\usepackage{setspace}
\usepackage{amsmath,amssymb,amsfonts,amsthm,mathtools}
\usepackage{bm}
\usepackage{booktabs}
\usepackage{multirow}
\usepackage{threeparttable}
\usepackage{array}
\usepackage{tabularx}
\usepackage{longtable}
\usepackage{graphicx}
\usepackage{adjustbox}
\usepackage{subcaption}
\usepackage{float}
\usepackage{caption}
\usepackage{enumitem}
\usepackage{hyperref}
\usepackage[nameinlink,capitalise,noabbrev]{cleveref}
\usepackage{url}
\usepackage{xcolor}
\usepackage{lscape}
\usepackage{pdflscape}
\usepackage{fancyhdr}
\usepackage{lastpage}
\usepackage{titlesec} 
\usepackage{microtype}
\usepackage{pifont}
\usepackage[
firstinits=true, 
useprefix=true, maxcitenames=3, maxbibnames=99, style=authoryear,
dashed=false,    
backend=biber, url=false, uniquename=false ]{biblatex}
\usepackage{tikz}
\usetikzlibrary{arrows.meta, positioning, shapes.geometric, fit, backgrounds}

\hypersetup{
    colorlinks=true,
    linkcolor=blue!60!black,
    citecolor=blue!60!black,
    urlcolor=blue!60!black
}

\newcommand{\Dirac}{\textsc{Dirac-3}}
\newcommand{\Gurobi}{\textsc{Gurobi}}
\newcommand{\SAC}{\textsc{SAC}}
\newcommand{\QUBO}{\textsc{QUBO}}
\newcommand{\HUBO}{\textsc{HUBO}}
\newcommand{\bone}{\beta_{1}}
\newcommand{\btwo}{\beta_{2}}
\newcommand{\CVaR}{\text{CVaR}_{5\%}}

\crefname{figure}{exhibit}{exhibits}
\Crefname{figure}{Exhibit}{Exhibits}
\crefname{table}{exhibit}{exhibits}
\Crefname{table}{Exhibit}{Exhibits}
\makeatletter
\let\c@table\c@figure

\makeatother

\titleformat{\section}{\normalfont\Large\bfseries}{}{0pt}{}
\titleformat{\subsection}{\normalfont\large\bfseries}{}{0pt}{}
\titleformat{\subsubsection}{\normalfont\normalsize\bfseries}{}{0pt}{}

\InputIfFileExists{results_ledger.tex}{}{}
\InputIfFileExists{narrative_stats.tex}{}{}

\newcommand{\InputTableOrPlaceholder}[1]{%
    \IfFileExists{#1}{%
        \resizebox{\linewidth}{!}{\input{#1}}%
    }{%
        \fbox{\parbox{0.92\linewidth}{\centering
        \textit{Missing generated exhibit file}\\
        \texttt{\detokenize{#1}}}}%
    }%
}

\newcommand{\IncludeFigureOrPlaceholder}[2][]{%
    \IfFileExists{#2}{%
        \includegraphics[#1]{#2}%
    }{%
        \fbox{\parbox[c][0.26\textheight][c]{0.92\linewidth}{\centering
        \textit{Missing generated exhibit file}\\
        \texttt{\detokenize{#2}}}}%
    }%
}

\title{Photonic Quantum Computing \emph{vs.} Classical Solvers\\
    in Constrained Factor Portfolio Optimization}

  \author{Nirvik Sahoo\thanks{\scriptsize Nirvik Sahoo is a research
      associate in the School of Computing and Information Systems at
      Singapore Management University in Singapore.  80 Stamford Road,
      Singapore 178902.  Email:
      \href{mailto:nirviks@suss.edu.sg}\texttt{nirviks@smu.edu.sg}
      } \and Chyng Wen Tee\thanks{\scriptsize Chyng Wen Tee is
      an associate professor in the Lee Kong Chian School of Business
      at Singapore Management University in Singapore. 50 Stamford
      Road, Singapore 178899. Email:
      \href{mailto:cwtee@smu.edu.sg}\texttt{cwtee@smu.edu.sg}}
    \and Paul Griffin \thanks{\scriptsize Paul Griffin is an associate
      professor in the School of Computing and Information Systems at
      Singapore Management University in Singapore.  80 Stamford Road,
      Singapore 178902.  Email:
      \href{mailto:paulgriffin@suss.edu.sg}\texttt{paulgriffin@smu.edu.sg}}}

\date{\today}
\begin{document}
\maketitle

Key Findings
\begin{itemize}
\item[\ding{111}] \underline{\it Narrow Quantum Advantage Range} The
  \Dirac{} entropy photonic annealer achieves superior risk-adjusted
  performance, delivering a peak Sharpe ratio of 0.76 and Calmar ratio
  of 0.567 compared to classical benchmarks. Yet this performance is
  highly localized within a narrow hyperparameter calibration range.
\item[\ding{111}] \underline{\it Robustness \& Tail Risk Stability}
  Mixed-integer programming (\Gurobi{}) consistently delivers the most
  stable tail-risk protection (CVaR = --0.863\%) and portfolio
  diversification, maintaining consistently low concentration across
  all market regimes without catastrophic failure modes.
  
\item[\ding{111}] \underline{\it Reinforcement Learning
    Vulnerabilities} The Soft Actor-Critic (\SAC{}) policy-gradient
  approach exhibits severe concentration failure under aggressive
  higher-moment penalties, highlighting the risks of unanchored
  policy-gradient methods in asset allocation.
\end{itemize}

\bigskip

Abstract\\

The authors present a rigorous empirical evaluation of three distinct
optimization paradigms for institutional factor portfolio
construction: an entropy-based photonic quantum annealer (\Dirac{},
Quantum Computing Inc.), a commercial mixed-integer programming solver
(\Gurobi{}), and a model-free deep reinforcement learning agent
(\SAC{}). Evaluating these pipelines on the Jensen--Kelly--Pedersen
13-factor equity library across 164 months test window, we implement a
full factorial penalty sweep comprising 48 hyperparameter
configurations that govern return, volatility, and skewness
trade-offs. Our findings demonstrate that while photonic hardware can
locate superior risk-return topologies within a narrow operating
range, classical mixed-integer programming remains superior for
risk-constrained mandates requiring tight tail-risk control and
cross-seed stability. Furthermore, we document structural failure
modes in reinforcement learning factor allocators under unanchored
higher-moment shaping. We translate these empirical results into
actionable, mandate-specific guidelines for quantitative portfolio
managers deploying advanced optimization engines.

\clearpage

\doublespacing

\section{Introduction}\label{sec:intro}

The intersection of quantitative asset management and advanced
computation has accelerated dramatically as portfolio managers seek to
solve increasingly complex, non-convex, and combinatorial portfolio
optimization problems. Recent surveys include \textcite{Orus2019,
  Herman2023}. Institutional asset allocation frequently incorporates
cardinal constraints, including limits on active factor bets, turnover
restrictions, transaction cost penalties, and higher-moment risk
shaping, that transform standard mean-variance formulations into
NP-hard mixed-integer problems, as detailed in
\textcite{Cornuejols2006, BertsimasFarrokhnia2023}.

To navigate these non-parametric search spaces, quantitative
researchers are exploring non-classical computing architectures,
notably quantum annealers and photonic quantum processors (see
\textcite{Mugel2022}). Concurrently, deep reinforcement learning (DRL)
has emerged as a dynamic approach to adaptive asset allocation, with
recent results documented by \textcite{Jiang2017, Ye2020}. However,
despite considerable academic enthusiasm, rigorous, multi-metric
head-to-head empirical evaluations comparing quantum hardware,
commercial classical solvers, and deep reinforcement learning under
identical financial constraints and factor universes remain scarce.

This paper addresses this gap by conducting a systematic empirical
study of three architecturally distinct optimization pipelines
evaluated on the Jensen--Kelly--Pedersen (JKP) 13-factor equity
library over a 164-month live test window by
\textcite{Jensen2023}. The three architectures evaluated are: (1)
\Dirac{} (Quantum Computing Inc.): an entropy-based photonic quantum
annealer that natively processes higher-order unconstrained binary
optimization (\HUBO{}) and continuous-variable objective functions
without the limitations of qubit encoding (see \textcite{QCI2024});
(2) \Gurobi{}: a commercial mixed-integer programming (MIP) solver
operating via branch-and-bound algorithms, serving as our
deterministic classical benchmark (see \textcite{Gurobi2024}); and (3)
\SAC{} (Soft Actor-Critic): an off-policy deep reinforcement learning
agent designed to learn continuous portfolio weight policies directly
from environmental rewards introduced by \textcite{Haarnoja2018}, and
subsequently adapted for long-only allocation by
\textcite{Almahdi2017, Jiang2017}.

To robustly assess how these solvers manage operational constraints
and investor risk/reward preferences, we implement a comprehensive
two-stage experimental design across 48 parameter configurations: a
primary sweep over volatility-penalization intensity ($\bone$) and a
joint sweep over volatility-penalization and skewness-shaping
($\btwo$).  Our empirical findings challenge common uncritical
assumptions regarding quantum supremacy or reinforcement learning
dominance in quantitative asset management.  While \Dirac{} captures
the global optimum in Sharpe ratio (0.760) and Calmar ratio (0.567),
its outperformance is restricted to a narrow hyperparameter
range. Conversely, \Gurobi{} provides superior tail-risk control (CVaR
= --0.863\%) and maintain a diversified holding across all parameter
configurations. Finally, we identify a serious structural
vulnerability in the \SAC{} pipeline: under unanchored secondary
moment penalties ($\btwo \geq 10$ with $\bone = 0$), the RL agent
suffers from severe portfolio concentration (HHI expanding to 0.593),
leading to deep drawdowns due to a lack of diversification.

This paper makes three distinct contributions to the quantitative
finance literature. First, we conduct a comprehensive, three-way
empirical evaluation comparing an entropy-based photonic quantum
annealer, a commercial branch-and-bound mixed-integer programming
solver, and a model-free deep reinforcement learning agent over an
established institutional equity factor library. By implementing a
granular two-stage penalty sweep, we map the complete risk-return and
higher-moment performance surface for each architecture. Second, we
uncover critical structural concentration dynamics inherent to each
solver class---most notably, the catastrophic factor concentration and
drawdown collapse exhibited by Soft Actor-Critic agents under
unanchored higher-moment shaping. These findings provide vital
insights into model fragility independent of raw return
maximization. Third, we translate these empirical insights into
actionable, mandate-specific guidelines for quantitative portfolio
managers and risk allocators deploying advanced computational
optimization engines in production environments.

The remainder of this article is organised as
follows. Section~\ref{sec:related} provides a brief reviews of recent
literature. Section~\ref{sec:framework} formalizes the mathematical
framework of the constrained optimization problem, furnishes the
economic interpretation of the hyperparameters along with the tuning
procedure, and outlines the data-to-QUBO pipeline.
Section~\ref{sec:results} details the dataset and performance metrics,
the empirical findings of the primary and joint hyperparameter sweeps,
along with numerical analyses on factor concentration
dynamics. Section~\ref{sec:discussion} discusses the practical
implications and recommendations based on our empirical observation,
and finally conclusions are drawn in Section~\ref{sec:conclusion}.

\section{Related Literature}\label{sec:related}

Modern quantitative portfolio construction frequently uses both
classical econometric models and machine learning paradigms. While
factor specifications historically relied on linear models, the
high-dimensional search space require algorithms capable of managing
non-linearities, non-normality, and higher order combinatorial
problems. The integration of traditional econometric approaches with
modern machine learning frameworks in portfolio management need not be
mutually exclusive.  In fact, recent research by \textcite{CF2020}
suggests that these two methodologies are complementary. However,
\textcite{LTKKF2023} has pointed out that interpretability remains an
ongoing issue. \textcite{KMZ2024} challenge the conventional wisdom
that simple, parsimonious models are superior for predicting financial
market returns, with both \textcite{LKKF2024} and \textcite{S2024}
highlighting the advantage of a more adaptive, data-driven approach
guided by practical considerations and empirical results.

In quantitative asset management, robust regularization is vital to
prevent overfitting of historical factor covariance matrices (see
\textcite{Ledoit2004, DeMiguel2009} for detail). While shrinkage
methodologies can address the estimation error in covariance inputs,
the calibration of Lagrange multipliers governing portfolio
constraints, such as turnover, volatility, and skewness, remains
largely heuristic. This paper systematically explores the
regularization surface defined by volatility penalties ($\bone$) and
skewness incentives ($\btwo$) across heterogeneous solver
technologies.

To handle complex constraints, classical portfolio optimization relies
heavily on exact branch-and-bound algorithms implemented via
commercial MIP solvers such as \Gurobi{} or CPLEX, which guarantee
global optimality within specified tolerances for cardinality- and
turnover-constrained problems (see \textcite{Cornuejols2006,
  BertsimasFarrokhnia2023}). At the same time, the quantitative
portfolio management space is also experiencing a paradigm shift from
static, single-period optimization to dynamic, sequential
decision-making under uncertainty (see \textcite{Kolm2025}). As
\textcite{Zhang2020} has highlighted, through the framing of portfolio
allocation as a Markov Decision Process (MDP), reinforcement learning
(RL) agents can interact with market environments to learn adaptive
policies directly from reward streams, naturally accommodating
multi-objective trade-offs, transaction costs, and feedback
effects. While classical quantitative approaches rely on strict
stationarity assumptions, RL frameworks excel in capturing temporal
dependencies and market frictions.

Nevertheless, \textcite{Bartram2021} has mentioned that institutional
adoption of RL is frequently hindered by data inefficiency, black-box
interpretability, and sensitivity to reward design . Among many
actor-critic methods, Soft Actor-Critic (\SAC{}) has been widely
adopted due to its entropy-regularized exploration and stability in
continuous action spaces (see \textcite{Haarnoja2018,
  Zarkias2019}). Practitioners often note that model-free RL agents
suffer from high seed sensitivity and can learn fragile policies under
poorly specified reward functions, a phenomenon we explicitly quantify
in our multi-moment penalty sweeps. Nevertheless, both
\textcite{Jiang2017, Ye2020} have identified that the dynamic asset
allocation field has been increasingly their leverage on deep
reinforcement learning technique more broadly.

To address these implementation hurdles, recent literature has
explored diverse architectures and hybrid paradigms. For instance,
\textcite{Han2023} bypasses the instability of raw weight generation
by utilizing deep deterministic policy gradients to dynamically
optimize risk budgeting parameters, while \textcite{Fjellavli2025}
advance multi-objective portfolio optimization by integrating return,
volatility, and liquidity objectives via Pareto
Q-learning. Comprehensive reviews by \textcite{Borkar2024} categorize
these underlying RL algorithms, ranging from Q-learning and deep
Q-networks to actor-critic methods, while broader surveys by
\textcite{Kolm2025} emphasize that RL is most successfully deployed as
a complementary decision engine that acts upon robust predictive
signals within simulated or data-rich environments.

Beyond classical and learning-based solvers, \textcite{Orus2019} posit
that the application of quantum-mechanical principles to financial
economics has evolved from theoretical foundations to empirical
validation. Quantum algorithms for derivative pricing, risk management
(specifically Value-at-Risk and Conditional Value-at-Risk via
amplitude estimation), and portfolio optimization have been
extensively studied by \textcite{Woerner2019, Herman2023}. In
portfolio theory, combinatorial selection problems are naturally
mapped to Quadratic Unconstrained Binary Optimization (\QUBO{}) or
Ising models. Early implementations on D-Wave quantum annealers by
\textcite{Mugel2022} demonstrated feasibility, but were frequently
constrained by synthetic datasets, small asset universes, or a lack of
direct comparison against commercial classical solvers under realistic
transaction costs. Our study bridges this gap by evaluating photonic
quantum hardware on a comprehensive equity factor library under full
transaction-cost and turnover penalties.

\section{Optimization Framework}\label{sec:framework}

We begin by presenting in full the mathematical framework behind our
portfolio optimization problem.  Let $\mathbf{w}_t \in \mathbb{R}^N$
denote the vector of portfolio weights across $N = 13$ equity factors
at rebalancing period $t$, subject to the budget constraint
$\sum_{i=1}^N w_{i,t} = 1$ and no-short-selling constraints
$w_{i,t} \ge 0$. The portfolio optimization agent seeks to maximize an
objective function that balances expected return against portfolio
volatility, higher-order return asymmetry (skewness), and portfolio
turnover:
\begin{equation}
  \max_{\mathbf{w}_t} \;
  \mathcal{R}(\mathbf{w}_t)
  \;=\;
  \underbrace{\mathbf{r}_t^\top \mathbf{w}_t}_{\text{expected return}}
  \;-\;
  \underbrace{\bone\, \widehat{\sigma}_{t,m}}_{\text{volatility penalty}}
  \;+\;
  \underbrace{\btwo\, \widehat{\kappa}_{t,m}}_{\text{skewness reward}}
  \;-\;
  \underbrace{c_{\text{tc}}\, \|\mathbf{w}_t - \mathbf{w}_{t-1}\|_1}_{\text{transaction cost penalty}},
  \label{eq:reward}
\end{equation}
where $\mathbf{r}_t$ is the realized factor return vector at month
$t$, $\widehat{\sigma}_{t,m}$ is the rolling $m$-period realized
volatility of the portfolio, $\widehat{\kappa}_{t,m}$ is the rolling
bias-corrected Fisher skewness coefficient of the portfolio return
series, and $c_{\text{tc}} = 0.002$ represents proportional
transaction costs.  The economic roles of the penalty hyperparameters
are defined as follows:
\begin{itemize}
\item Primary Volatility Penalty ($\bone$): Serves as the primary
  risk-aversion parameter. Setting $\bone = 0$ maximizes net return
  without explicit volatility aversion. As $\bone$ increases, the
  objective penalizes return volatility with increasing severity,
  forcing the optimizer toward lower-volatility factor combinations
  and dampening turnover.
  
\item Secondary Skewness Incentive ($\btwo$): Governs preferences for
  return distribution asymmetry. Positive values ($\btwo>0$) reward
  positive realized skewness (fat right tails, mitigating drawdown
  risk), while negative values penalize right-skewed distributions.
\end{itemize}

To map factor selection and weighting into a format natively
executable by the \Dirac{} photonic annealer and \Gurobi{}, we
formulate the mean-variance factor-subset selection problem as a
Quadratic Unconstrained Binary Optimization (\QUBO{}) problem.  Let
$\mathbf{x} \in \{0,1\}^N$ be a binary decision vector indicating
whether factor $i$ is included in the active portfolio
($\mathbf{x}_i = 1$) or excluded ($\mathbf{x}_i = 0$). The
optimization objective is expressed as:
\begin{equation}
  \min_{\mathbf{x} \in \{0,1\}^N} \;
  \mathbf{x}^\top Q\, \mathbf{x}
  \quad \text{where} \quad
  Q_{ij} =
  \begin{cases}
    -\alpha_i + c_{\text{pos}} + \lambda\,\Sigma_{ii} + \delta_i & i = j \\
    2\lambda\,\Sigma_{ij} & i \neq j,
  \end{cases}
  \label{eq:qubo}
\end{equation}

where $\boldsymbol{\alpha} \in \mathbb{R}^N$ is the expected return
vector generated by the policy model,
$\boldsymbol{\Sigma} \in \mathbb{R}^{N \times N}$ is the Ledoit--Wolf
shrinkage covariance matrix of factor returns, $c_{\text{pos}} = 0.01$
is a fixed cardinality-opening cost, $\lambda = 1.0$ is the baseline
risk-aversion scalar, and $\delta_i = \pm \tau_{\text{pen}}$
represents a dynamic turnover penalty that discourages style drift by
penalizing newly introduced factors and rewarding retained factors
from the prior period ($\tau_{\text{pen}} = 0.10$).

Once the optimal binary subset $\mathcal{S} = \{i : x_i^* = 1\}$ is
determined by the solver, continuous portfolio weights
$\mathbf{w}_{\mathcal{S}}$ are derived via restricted mean-variance
optimization over the selected factor subset, subject to individual
factor bounds $w_i \in [0.00, 0.60]$. Finally, an exponential
smoothing filter is applied to transition portfolio weights smoothly
between rebalancing dates:
\begin{equation}
  \mathbf{w}_t = (1 - \eta)\mathbf{w}_{t-1} + \eta \mathbf{w}_{\mathcal{S}},
\end{equation}
with the smoothing parameter $\eta = 0.10$.

The solution $\mathbf{x}^*$ identifies the included factor
subset. Given the selected subset $\mathcal{S} = \{i : x_i^* = 1\}$,
continuous weights are determined by the restricted mean-variance
allocation
$\mathbf{w}_{\mathcal{S}} = \boldsymbol{\Sigma}_{\mathcal{S}}^{-1}
\boldsymbol{\alpha}_{\mathcal{S}} \big/ \mathbf{1}^\top
\boldsymbol{\Sigma}_{\mathcal{S}}^{-1}
\boldsymbol{\alpha}_{\mathcal{S}}$, clipped to $[w_{\min}, w_{\max}]$
and renormalized. The final portfolio weights $\mathbf{w}$ apply an
exponential smoothing blend
$\mathbf{w}_t = (1 - \eta)\, \mathbf{w}_{t-1} + \eta\,
\mathbf{w}_{\mathcal{S}}$ with blend factor $\eta = 0.10$ to further
dampen turnover.

Two nested experimental designs are studied. The $\bone$ sweep fixes
$\btwo = 0$ and varies $\bone \in \{0, 0.5, 1, 2, 5, 10, 20, 50\}$ (8
configurations per pipeline, 24 total). The joint sweep varies both:
$\bone \in \{0, 1\}$ and $\btwo \in \{0, 0.5, 1, 2, 5, 10, 20, 50\}$
(16 configurations per pipeline, 48 total). Independent seeds are used
per configuration to estimate cross-seed variance: three for \Dirac{}
given the per-run computational cost of photonic hardware access, and
five for \Gurobi{} and \SAC{}. All Dirac-3 jobs use a relaxation
schedule of $\text{RS}=2$, selected to balance solution quality
against per-run wall-clock cost given the photonic hardware's
queue-based access model and the scale of the 48-configuration sweep.

\begin{figure}[!htb]
  \centering \resizebox{0.75\textwidth}{!}{%
    \begin{tikzpicture}[ box/.style={rectangle, draw, rounded
        corners=3pt, minimum width=3.2cm, minimum height=0.75cm,
        align=center, font=\small},
      arr/.style={-{Stealth[length=5pt]}, thick},
      branch/.style={rectangle, draw, dashed, rounded corners=3pt,
        minimum width=2.6cm, minimum height=0.65cm, align=center,
        font=\small\itshape} ]

      \node[box] (data) at (0,10) {JKP Factor
        Returns\\$\mathbf{R}\in\mathbb{R}^{T\times
          N}$}; \node[box] (cov) at (0,8.6) {Ledoit--Wolf\\Covariance
        $\hat{\Sigma}_t$}; \node[box] (policy) at (0,7.2) {Policy
        Network\\Action $\mathbf{a}_t$};

      \node[box] (qubo) at (-2,5.4) {\QUBO{}
        Construction\\(Eq.~\ref{eq:qubo})}; \node[branch] (sac) at
      (3,5.4) {\SAC{}\\(softmax on $\mathbf{a}_t$)};

      \node[box] (dispatch) at (-2,3.8) {Solver Dispatch};

      \node[branch] (dirac) at (-3.6,2.2) {\Dirac{}\\(photonic)};
      \node[branch] (gurobi) at (-0.4,2.2) {\Gurobi{}\\(MIP/B\&B)};

      \node[box] (subset) at (0,0.4) {Subset
        $\mathcal{S}^*$\,/\,Weights}; \node[box] (mv) at (0,-1.0)
      {Restricted MV Weights\\+ Smoothing}; \node[box] (reward) at
      (0,-2.4) {Portfolio Return $r_t$,\\Reward Eq.~\ref{eq:reward}};
      \node[box] (metrics) at (0,-3.8) {11 Performance Metrics};

      \draw[arr] (data) -- (cov); \draw[arr] (cov) -- (policy);
      \draw[arr] (policy.south) -- ++(0,-0.3) -| (qubo.north);
      \draw[arr] (policy.south) -- ++(0,-0.3) -| (sac.north);
      \draw[arr] (qubo) -- (dispatch); \draw[arr] (dispatch.south
      west) -- ++(-0.3,-0.3) -| (dirac.north); \draw[arr]
      (dispatch.south east) -- ++(0.3,-0.3) -| (gurobi.north);

      \draw[arr] (dirac.south) |- (subset.west); \draw[arr]
      (gurobi.south) -- ++(0,-0.4) -| (subset.north); \draw[arr]
      (sac.south) |- (subset.east);

      \draw[arr] (subset) -- (mv); \draw[arr] (mv) -- (reward);
      \draw[arr] (reward) -- (metrics);

      \draw[arr] (reward.east) -- ++(5.5,0) |- node[pos=0.25, right,
      font=\small] {Update policy} (policy.east);

    \end{tikzpicture}
  }
  \caption{End-to-end data-to-portfolio pipeline. All three solvers
    share Stages 1--3 (data ingestion, covariance estimation, and
    policy action). \Dirac{} and \Gurobi{} proceed through Stage 4
    (\QUBO{} construction) and Stage 5 (solver dispatch): \Dirac{}
    submits the polynomial to the photonic annealer API, while
    \Gurobi{} solves the binary quadratic program via
    branch-and-bound. \SAC{} bypasses both the \QUBO{} construction
    and dispatch stages, applying a softmax directly to the raw policy
    action vector $\mathbf{a}_t$. Stages 6 onward (weight
    reconstruction, reward computation, metric aggregation) are
    shared.}
  \label{fig:pipeline}
\end{figure}
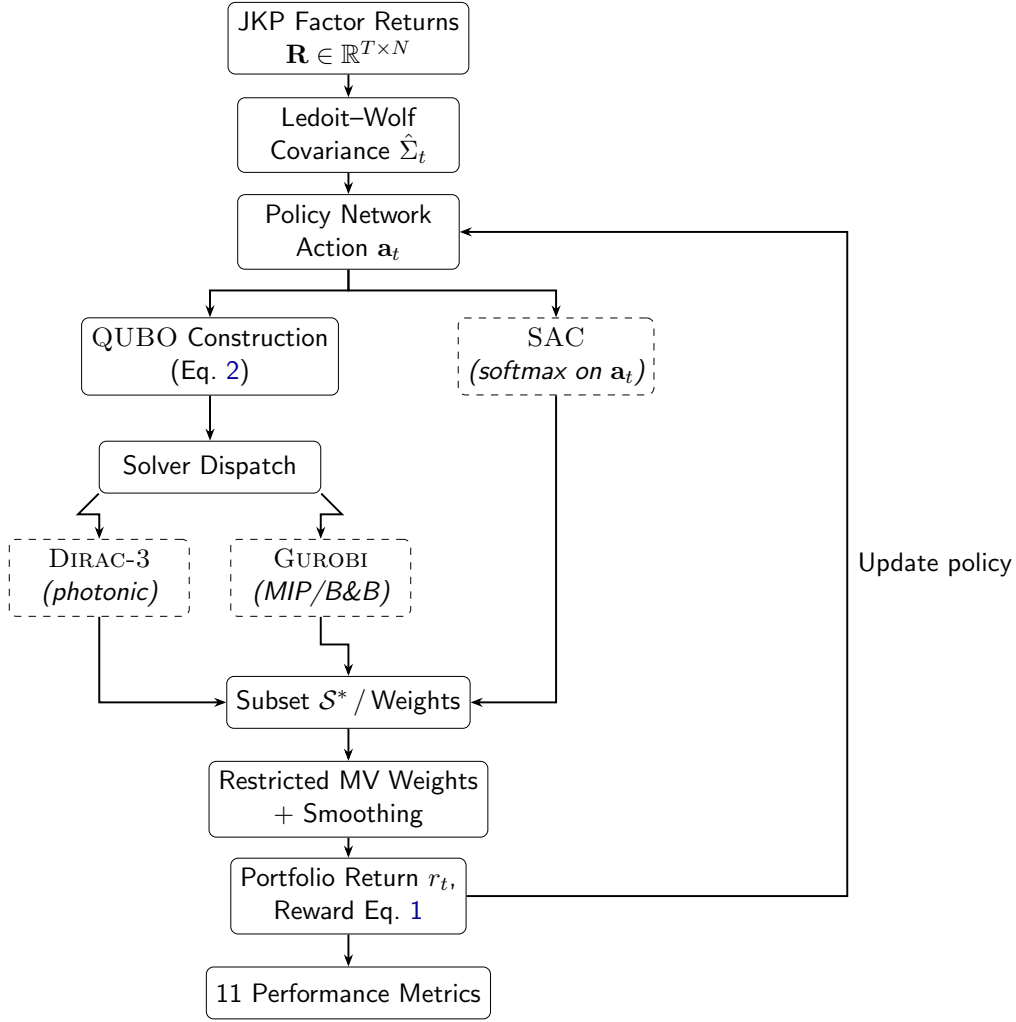

Next, we proceed to describe in full the end-to-end computational
pipeline through which raw monthly factor returns are transformed into
a \QUBO{} problem, dispatched to each solver, and converted back to
portfolio weights. The three solvers share a common
problem-construction stage and diverge only at the solver-dispatch
step. Exhibit~\ref{fig:pipeline} provides an overview of the portfolio
construction process, illustrating our six-stage architecture.  To
ensure fair comparisons, the experimental architecture executes across
a six-stage data-to-portfolio pipeline that standardizes inputs across
all three solver engines.  The process begins with data ingestion and
preprocessing, where monthly factor returns from the JKP library are
filtered to isolate $N=13$ value-weighted, capped US equity anomaly
factors spanning a historical panel of $T=164$ months
($\mathbf{R} \in \mathbb{R}^{T \times N}$).  The dataset is
partitioned chronologically using an 80/20 train-test split, reserving
131 months for model training and 33 months for out-of-sample
evaluation. All reported performance metrics are computed exclusively
over the test window. Following ingestion, the framework performs
rolling covariance estimation at each rebalancing date $t$.  To handle
the relatively short 24-month lookback window ($m_{\text{cov}} = 24$)
relative to the factor dimension, the covariance matrix
$\hat{\boldsymbol{\Sigma}}_t$ is estimated using the shrinkage
approach introduced by \textcite{Ledoit2004}. The matrix is
subsequently symmetrized via
$\hat{\boldsymbol{\Sigma}}_t \leftarrow
\frac{1}{2}(\hat{\boldsymbol{\Sigma}}_t +
\hat{\boldsymbol{\Sigma}}_t^\top) + \varepsilon\mathbf{I}$ with a
small diagonal perturbation ($\varepsilon = 10^{-8}$) to ensure
numerical positive definiteness and prevent conditioning instability.

The next two stages cover signal generation and problem formulation,
where a fundamental structural distinction emerges between the pure
reinforcement learning architecture and the solver-augmented hybrid
pipelines. Under the standalone Soft Actor-Critic (\SAC{}) agent, the
policy network outputs a continuous action vector
$\mathbf{a}_t \in \mathbb{R}^N$ at each step, directly parametrizing
the portfolio weights through a softmax transformation
($\mathbf{w}_t = \text{softmax}(\mathbf{a}_t)$) and calculating
rewards directly from realized portfolio returns without formulating
an intermediate optimization problem. Conversely, for \Dirac{} and
\Gurobi{}, the policy network acts as a signal generator whose
continuous action vector represents an expected return forecast
($\boldsymbol{\alpha} = \mathbf{a}_t$). This expected return signal,
along with the shrinkage covariance matrix
$\hat{\boldsymbol{\Sigma}}_t$ and the preceding binary allocation
vector $\mathbf{x}^{\text{prev}}$, is assembled into a sparse
Quadratic Unconstrained Binary Optimization (\QUBO{}) matrix $Q$. The
resulting polynomial representation stores linear diagonal terms
$Q_{ii} = -\alpha_i + c_{\text{pos}} + \lambda\Sigma_{ii} + \delta_i$
to capture return targets, position costs, individual variance, and
turnover penalties, alongside non-zero quadratic off-diagonal terms
$Q_{ij} = 2\lambda\Sigma_{ij}$ to model pairwise factor covariances.

In the execution phase, the formulated objective is dispatched
according to each architecture's specific solver interface. For the
photonic quantum pipeline, the sparse polynomial is submitted to the
\Dirac{} photonic annealer, where returning candidate binary vectors
from which the state yielding the lowest objective energy
$\mathbf{x}^*$ is extracted. For the classical benchmark, the problem
is formulated as a binary quadratic program and solved via
branch-and-bound, thresholding the binary solution at
$x_i^* = \mathbb{I}[x_i > 0.5]$. The pure \SAC{} agent bypasses this
dispatch phase entirely, relying on its internal policy network to
emit direct weight updates evaluated over a 6-month rolling window for
volatility and skewness penalties.

Finally, the pipeline concludes with weight reconstruction and
portfolio evaluation. Using the binary factor subset
$\mathcal{S} = \{i : x_i^* = 1\}$ identified by the solver, continuous
weights $\mathbf{w}_{\mathcal{S}}$ are derived via restricted
mean-variance optimization over the selected factors, subject to box
constraints $[w_{\min}, w_{\max}] = [0.00, 0.60]$ and unit sum
normalization. To dampen excessive trading activity, an exponential
smoothing filter blends the newly optimized subset weights with prior
allocations according to
$\mathbf{w}_t = (1-\eta)\mathbf{w}_{t-1} +
\eta\mathbf{w}_{\mathcal{S}}$ using a smoothing factor of
$\eta = 0.10$. These smoothed final weights are subsequently used to
record realized portfolio returns
$r_{\text{port},t} = \mathbf{w}_t^\top \mathbf{r}_t$, compute
fractional $L_1$-norm turnover
$\text{TO}_t = \Vert{}\mathbf{w}_t - \mathbf{w}_{t-1}\Vert{}_1$, and
supply state observations to feed the policy network for the
subsequent step. A detailed schematic of the full pipeline is
summarized in Exhibit~\ref{fig:pipeline}.

\section{Empirical Analyses}
\label{sec:results}

\subsection{Data and Experimental Design}

We use the Jensen--Kelly--Pedersen global factor library
by \textcite{Jensen2023}, restricting to 13 US equity anomaly factors:
value, momentum, short-term reversal, quality, profitability,
investment, low-risk, low-leverage, seasonality, accruals, debt
issuance, idiosyncratic volatility, and size.  Monthly factor returns
span 164 months (approximately 13.7 years) and cover a full market
cycle including the 2020 COVID-19 drawdown and the 2022
rate-tightening episode. Returns are net of transaction costs imputed
at the level of \textcite{Jensen2023}. The factor universe is filtered
to include only value-weighted, capped return series with no missing
observations across the full panel.

To evaluate pipeline performance, we assess eleven financial metrics,
categorized into return, risk-adjusted, tail-risk, and
portfolio-composition measures.  For returns and risk, we measure
annual return (\%), and annual volatility (\%), both are annualized
scaled from the mean monthly raw metrics. Risk-adjusted performance is
evaluated using Sharpe Ratio (our primary metric) and Sortino Ratio,
which replaces total volatility with downside deviation
$\sigma_{\text{down}}$ computed exclusively over negative returns.

Tail risks and distribution shapes are captured via realized skewness
(the Fisher bias-corrected third standardised moment of monthly
returns), conditional value-at-risk ($\CVaR$, \%) expressed as the
mean return over the worst 5\% of observations, and max drawdown (\%,
MDD), defined as the peak-to-trough percentage decline over the
evaluation window. We also use Calmar Ratio which combines return and
drawdown by dividing annualized return by the absolute maximum
drawdown.

Finally, we track portfolio structure and execution friction through
average turnover, measured as the mean $L_1$-norm of weight changes
$\|\mathbf{w}_t - \mathbf{w}_{t-1}\|_1$ between consecutive
months. Concentration is quantified using the Herfindahl–Hirschman
Index (HHI), calculated as $\sum_{i=1}^N w_i^2$, where a uniform
allocation yields $1/N \approx 0.077$ for $N=13$, and the CR-5
concentration ratio, defined as the sum of the top-5 factor weights.

\subsection{Results Overview}

Exhibit~\ref{tab:beta1_summary} reports peak metrics per pipeline
across the $\bone$ sweep ($\btwo = 0$ throughout). The results reveal
differentiated performance profiles that vary by metric and
configuration. No single pipeline dominates on all metrics
simultaneously: \Dirac{} leads on Calmar ratio and maximum drawdown at
its optimal settings, \Gurobi{} leads on Sharpe ratio and $\CVaR$, and
\SAC{} records the best realized skewness at $\bone = 0$ and the
lowest HHI at high $\bone$.

\begin{table}[htbp]
  \centering
  \caption{Peak performance across the $\bone$ sweep ($\btwo = 0$).}
  \label{tab:beta1_summary}
  \small
  \begin{tabular}{@{}lllll@{}}
    \toprule
    Metric         & \SAC{}                    & \Gurobi{}                 & \Dirac{}                  & Winner \\
    \midrule
    Peak Sharpe    & 0.572 ($\bone\!=\!50$)    & 0.718 ($\bone\!=\!5$)     & 0.715 ($\bone\!=\!2$)     & \Gurobi{} \\
    Best Calmar    & 0.214 ($\bone\!=\!10$)    & 0.362 ($\bone\!=\!5$)     & 0.376 ($\bone\!=\!1$)     & \Dirac{} \\
    Best $\CVaR$   & --1.223 ($\bone\!=\!50$)  & --0.983 ($\bone\!=\!0$)   & --1.015 ($\bone\!=\!1$)   & \Gurobi{} \\
    Best Skewness  & 0.332 ($\bone\!=\!0$)     & 0.287 ($\bone\!=\!50$)    & 0.265 ($\bone\!=\!2$)     & \SAC{} \\
    Lowest HHI     & 0.113 ($\bone\!=\!20$)    & 0.136 ($\bone\!=\!1$)     & 0.192 ($\bone\!=\!0$)     & \SAC{} \\
    Lowest MDD     & --5.99\%                  & --4.33\% ($\bone\!=\!5$)  & --3.37\% ($\bone\!=\!1$)  & \Dirac{} \\
    \bottomrule
  \end{tabular}
\end{table}

Exhibit~\ref{tab:combined_beta1} presents a granular, head-to-head
performance comparison of the three execution pipelines across the
primary penalty parameter grid
($\beta_1 \in \{0, 1, 2, 5, 10, 20, 50\}$ with $\beta_2 = 0$). This
sweep allows us to trace the sensitivity of risk-adjusted returns,
tail-risk profiles, and portfolio concentration to hyperparameter
selection across distinct algorithmic paradigms: quantum-annealing
heuristics (\Dirac{}), exact mixed-integer linear programming
(\Gurobi{}), and stochastic reinforcement learning (\SAC{}).

\begin{table}[!htb]
  \centering \scriptsize
  \caption{All-pipeline comparison across
    $\bone \in \{0, 1, 2, 5, 10, 20, 50\}$ at $\btwo = 0$. }
  \label{tab:combined_beta1}
  \begin{tabular}{@{}lccccccc@{}}
    \toprule
    Metric        & $\bone=0$  & $\bone=1$  & $\bone=2$  & $\bone=5$  & $\bone=10$  & $\bone=20$  & $\bone=50$ \\
    \midrule
    \multicolumn{8}{l}{\emph{Ann.\ Return (\%)}} \\
    \quad SAC     & 1.28       & 1.20       & 1.85       & 1.46       & 1.39        & 1.45        & 1.05 \\
    \quad Gurobi  & 0.92       & 0.98       & 1.34       & 1.29       & 1.25        & 1.34        & 1.12 \\
    \quad Dirac   & 1.27       & 1.29       & 1.46       & 1.43       & 1.67        & 1.51        & 1.20 \\
    \addlinespace
    \multicolumn{8}{l}{\emph{Ann.\ Volatility (\%)}} \\
    \quad SAC     & 2.64       & 2.25       & 3.67       & 2.95       & 2.49        & 2.71        & 1.88 \\
    \quad Gurobi  & 1.68       & 1.77       & 2.14       & 1.81       & 2.18        & 2.07        & 1.86 \\
    \quad Dirac   & 1.86       & 1.86       & 2.05       & 2.51       & 2.82        & 2.23        & 1.86 \\
    \addlinespace
    \multicolumn{8}{l}{\emph{Sharpe ($\sqrt{12}$)}} \\
    \quad SAC     & 0.492      & 0.539      & 0.506      & 0.505      & 0.571       & 0.550       & 0.572 \\
    \quad Gurobi  & 0.563      & 0.568      & 0.629      & 0.718      & 0.616       & 0.621       & 0.612 \\
    \quad Dirac   & 0.630      & 0.697      & 0.715      & 0.551      & 0.610       & 0.692       & 0.652 \\
    \addlinespace
    \multicolumn{8}{l}{\emph{Sortino}} \\
    \quad SAC     & 0.536      & 0.561      & 0.544      & 0.530      & 0.603       & 0.573       & 0.598 \\
    \quad Gurobi  & 0.600      & 0.600      & 0.671      & 0.764      & 0.643       & 0.659       & 0.638 \\
    \quad Dirac   & 0.672      & 0.731      & 0.762      & 0.587      & 0.644       & 0.731       & 0.693 \\
    \addlinespace
    \multicolumn{8}{l}{\emph{Realized Skewness}} \\
    \quad SAC     & 0.332      & 0.014      & --0.237    & 0.049      & 0.006       & --0.131     & --0.377 \\
    \quad Gurobi  & 0.086      & 0.037      & --0.434    & 0.159      & 0.197       & 0.013       & 0.287 \\
    \quad Dirac   & --0.157    & 0.263      & 0.265      & 0.016      & --0.075     & --0.308     & 0.014 \\
    \addlinespace
    \multicolumn{8}{l}{\emph{$\CVaR$ (\%)}} \\
    \quad SAC     & --1.527    & --1.335    & --2.388    & --1.991    & --1.521     & --1.690     & --1.223 \\
    \quad Gurobi  & --0.983    & --1.042    & --1.334    & --1.049    & --1.154     & --1.248     & --1.052 \\
    \quad Dirac   & --1.221    & --1.015    & --1.176    & --1.578    & --1.678     & --1.355     & --1.201 \\
    \addlinespace
    \multicolumn{8}{l}{\emph{Max Drawdown (\%)}} \\
    \quad SAC     & --9.45     & --6.50     & --12.23    & --10.35    & --7.23      & --7.96      & --5.99 \\
    \quad Gurobi  & --5.77     & --5.32     & --6.14     & --4.33     & --6.67      & --6.69      & --5.63 \\
    \quad Dirac   & --4.48     & --3.37     & --6.48     & --8.22     & --7.79      & --4.85      & --6.00 \\
    \addlinespace
    \multicolumn{8}{l}{\emph{Calmar Ratio}} \\
    \quad SAC     & 0.143      & 0.205      & 0.181      & 0.167      & 0.214       & 0.190       & 0.200 \\
    \quad Gurobi  & 0.181      & 0.205      & 0.271      & 0.362      & 0.205       & 0.224       & 0.224 \\
    \quad Dirac   & 0.305      & 0.376      & 0.333      & 0.186      & 0.238       & 0.352       & 0.286 \\
    \addlinespace
    \multicolumn{8}{l}{\emph{Avg.\ Turnover}} \\
    \quad SAC     & 0.021      & 0.021      & 0.020      & 0.020      & 0.021       & 0.021       & 0.019 \\
    \quad Gurobi  & 0.021      & 0.025      & 0.025      & 0.018      & 0.024       & 0.018       & 0.023 \\
    \quad Dirac   & 0.017      & 0.021      & 0.024      & 0.022      & 0.021       & 0.028       & 0.019 \\
    \addlinespace
    \multicolumn{8}{l}{\emph{HHI}} \\
    \quad SAC     & 0.138      & 0.146      & 0.229      & 0.172      & 0.119       & 0.113       & 0.141 \\
    \quad Gurobi  & 0.139      & 0.136      & 0.156      & 0.157      & 0.141       & 0.170       & 0.137 \\
    \quad Dirac   & 0.192      & 0.218      & 0.210      & 0.225      & 0.212       & 0.240       & 0.245 \\
    \addlinespace
    \multicolumn{8}{l}{\emph{CR-5}} \\
    \quad SAC     & 0.686      & 0.725      & 0.723      & 0.724      & 0.662       & 0.642       & 0.707 \\
    \quad Gurobi  & 0.719      & 0.715      & 0.821      & 0.816      & 0.735       & 0.782       & 0.674 \\
    \quad Dirac   & 0.860      & 0.834      & 0.898      & 0.929      & 0.965       & 0.958       & 0.945 \\
    \bottomrule
  \end{tabular}%
\end{table}

\Dirac{} demonstrates a highly localized optimal operating window
concentrated at $\beta_1 \in \{1, 2\}$. At $\beta_1 = 1$, \Dirac{}
attains the sweep's premier Calmar ratio of $0.376$, accompanied by a
maximum drawdown (MDD) of --3.37\% and a robust annualized Sharpe
ratio of 0.697. Moving to $\beta_1 = 2$, \Dirac{} reaches its peak
Sharpe ratio of 0.715, the highest achieved by any pipeline at this
parameter setting, while maintaining a competitive annual volatility
of 2.05\% and positive realized skewness (+0.265). Factor attribution
reveals that this optimal window is structurally anchored by
quality-adjacent, persistent-signal factors, specifically accruals
(20.5\%), investment (14.5\%), and low-leverage (10.2\%). This
structural tilt directly underpins the pipeline's superior tail-risk
attenuation.

Conversely, \Gurobi{} achieves its apex performance at $\beta_1 = 5$,
posting a Sharpe ratio of 0.718, a Sortino ratio of 0.764, and a muted
annualized volatility of 1.81\%. The economic mechanism driving
\Gurobi{}'s success at this setting is its exceptionally balanced
factor distribution, highlighted by a Herfindahl-Hirschman Index (HHI)
of 0.157 and a top-5 concentration ratio (CR-5) of 0.816. Low-leverage
(13.9\%) and debt-issuance (10.0\%) serve as co-leaders within a
broadly diversified multi-factor matrix, reflecting the deterministic
convergence guarantees inherent to the branch-and-bound solver.

While \SAC{} generates the highest nominal annualized return in the
sweep (1.85\% at $\beta_1 = 2$), this return is achieved at the
expense of severe tail risk exposure and volatility expansion. The
excess return over \Gurobi{} at $\beta_1 = 2$ (1.85\% vs. 1.34\%)
fails to compensate for a fourfold expansion in Conditional
Value-at-Risk ($\CVaR$ reaching --2.388\%), a negative realized
skewness (--0.237), and a significant volatility premium relative to
the exact solver.

For institutional mandates operating under strict drawdown
constraints, tail-risk metrics ($\CVaR$ and MDD) are
paramount. \Gurobi{} exhibits the most stable downside profile across
the board. At $\beta_1 = 0$, \Gurobi{} records the sweep's lowest
$\CVaR$ of --0.983\%, supported by a tight compositional bandwidth
where HHI remains bound between 0.136 and 0.170 across all $\beta_1$
specifications.

\Dirac{} displays exceptional tail risk protection at $\beta_1 = 1$
($\CVaR$ of --1.015\% and MDD of --3.37\%). However, its downside
metrics deteriorate rapidly outside its optimal window, particularly
as portfolio concentration increases. At high penalty thresholds
($\beta_1 \in \{10, 20, 50\}$), \Dirac{}'s CR-5 climbs to extreme
levels (peaking at 0.965 at $\beta_1 = 10$), indicating an
over-concentration in a small subset of factors that erodes the
benefit of diversification without yielding proportional gains in
Sharpe or Calmar ratios.

\SAC{} consistently trails in downside metrics. Although its $\CVaR$
improves to --1.223\% at $\beta_1 = 50$, confirming that the
stochastic actor benefits from strong penalty regularization, its
corresponding Sharpe (0.572) and Calmar (0.200) ratios remain
sub-optimal compared to the deterministic and heuristic baselines.

Beyond point-estimate performance, reproducibility is a critical
criterion for production deployment in quantitative finance. \Gurobi{}
and \Dirac{} exhibit strong cross-seed stability. For instance, at
$\beta_1 = 20$, \Gurobi{} achieves a Calmar ratio of 0.224 with a
cross-seed return standard deviation of just 0.10 percentage points,
and its cross-seed Sharpe standard deviation remains strictly below
0.05. \Dirac{} maintains a similarly controlled cross-seed Sharpe
standard deviation ($<$0.08) across its entire $\beta_1$ grid.

In stark contrast, \SAC{} suffers from pronounced seed instability.
At $\beta_1 = 20$, the cross-seed standard deviation of \SAC{}'s
Sharpe ratio reaches 0.206, spanning a wide range from 0.168 to 0.718
(a spread of 0.55 points). This high variance introduces substantial
model risk, rendering unconstrained reinforcement learning pipelines
difficult to calibrate for live production without extensive ensemble
regularization.

The comprehensive empirical evidence above indicates a clear
bifurcation in pipeline utility: \Gurobi{} is best suited for
institutional mandates requiring deterministic feasibility
certificates, low turnover stability, and predictable drawdown
boundaries, with $\beta_1 = 5$ and $\beta_1 = 20$ representing the
optimal operational setpoints.  \Dirac{} offers superior risk-adjusted
alpha capture via its quantum-annealing heuristics, provided
hyperparameter selection is strictly restricted to the narrow
$\beta_1 \in \{1, 2\}$ sweet spot to avoid hyper-concentration.
\SAC{} captures high absolute momentum returns but requires enhanced
regularization and ensemble smoothing to overcome the seed-dependent
variance that undermines its out-of-sample reliability.

\subsection{Detailed Hyperparameter Sweep}

Exhibit~\ref{tab:joint_summary} reports peak metrics across the joint
sweep. The introduction of the $\btwo$ (skewness incentive) leads to
substantially different performance across the three pipelines, and
reveals both the global optimum configurations for each solver and the
failure modes that arise under large penalty magnitudes.

\begin{table}[htbp]
  \centering
  \caption{Peak performance across the joint $(\bone, \btwo)$ sweep. }
  \label{tab:joint_summary}
  \small \setlength{\tabcolsep}{4pt}
  \begin{tabular}{@{}lllll@{}}
    \toprule
    Metric         & \SAC{}              & \Gurobi{}             & \Dirac{}               & Winner \\
    \midrule
    Peak Sharpe    & 0.602 at $(1,0.5)$  & 0.715 at $(0,50)$     & 0.760 at $(0,1)$       & \Dirac{} \\
    Best Calmar    & 0.262 at $(0,0.5)$  & 0.324 at $(1,50)$     & 0.567 at $(0,1)$       & \Dirac{} \\
    Best $\CVaR$   & --1.228 at $(0,2)$  & --0.863 at $(0,1)$    & --0.895 at $(1,10)$    & \Gurobi{} \\
    Best Skewness  & 0.475 at $(1,20)$   & 0.377 at $(1,5)$      & 0.597 at $(0,50)$      & \Dirac{} \\
    Lowest HHI     & 0.132 at $(1,0.5)$  & 0.135 at $(0,0.5)$    & 0.161 at $(1,0.5)$     & \Gurobi{} \\
    Lowest MDD     & --5.99\%            & --4.95\% at $(0,50)$  & --2.63\% at $(0,0.5)$  & \Dirac{} \\
    \bottomrule
  \end{tabular}%
\end{table}


Exhibit~\ref{tab:joint_beta0_beta1} provide full per-configuration
results for $\bone = 0$ and $\bone = 1$, respectively.  \Dirac{}'s
global optimum in the full experiment is $(\bone=0, \btwo=1)$: Sharpe
0.760, Sortino 0.841, Calmar 0.567, MDD --3.47\%, and $\CVaR$
--1.278\%.  Factor composition at this setting is highly concentrated:
accruals (25.9\%) and investment (19.6\%) jointly account for 45.5\%
of total portfolio weight.  This concentration in quality-adjacent
factors is associated with superior skewness and $\CVaR$ outcomes
relative to failure-mode configurations where momentum or low-risk
dominate.

\begin{table}[!htb]
  \scriptsize \centering
  \caption{All-pipeline comparison across $\btwo$ at $\bone = 0$ and
    $\bone = 1$. All figures are seed means.}
  \label{tab:joint_beta0_beta1}
  \small \setlength{\tabcolsep}{3pt}
  \begin{tabular}{@{}lc|ccc|ccc|ccc@{}}
    \toprule
    & & \multicolumn{3}{c|}{Sharpe} & \multicolumn{3}{c|}{$\CVaR$ (\%)} & \multicolumn{3}{c}{HHI} \\
    $\bone$  & $\btwo$  & \SAC{}  & \Gurobi{}  & \Dirac{}  & \SAC{}   & \Gurobi{}  & \Dirac{}  & \SAC{}  & \Gurobi{}  & \Dirac{} \\
    \midrule
    \multirow{8}{*}{0} 
    & 0    & 0.536  & 0.603  & 0.646  & --1.933  & --1.115  & --1.419  & 0.201  & 0.148  & 0.353 \\
    & 0.5  & 0.521  & 0.507  & 0.721  & --2.145  & --1.135  & --1.107  & 0.154  & 0.135  & 0.196 \\
    & 1    & 0.453  & 0.565  & 0.760  & --2.233  & --0.863  & --1.278  & 0.177  & 0.163  & 0.340 \\
    & 2    & 0.485  & 0.590  & 0.590  & --1.228  & --1.081  & --1.517  & 0.154  & 0.157  & 0.268 \\
    & 5    & 0.436  & 0.568  & 0.720  & --1.832  & --1.440  & --0.979  & 0.200  & 0.140  & 0.235 \\
    & 10   & 0.304  & 0.605  & 0.666  & --2.331  & --1.360  & --1.904  & 0.333  & 0.170  & 0.245 \\
    & 20   & 0.088  & 0.617  & 0.421  & --2.237  & --1.230  & --1.255  & 0.593  & 0.140  & 0.281 \\
    & 50   & 0.163  & 0.715  & 0.575  & --2.066  & --0.911  & --1.255  & 0.520  & 0.187  & 0.225 \\
    \midrule
    \multirow{8}{*}{1} 
    & 0    & 0.555  & 0.563  & 0.575  & --1.339  & --1.092  & --1.157  & 0.153  & 0.138  & 0.181 \\
    & 0.5  & 0.602  & 0.566  & 0.653  & --1.363  & --1.099  & --1.099  & 0.132  & 0.163  & 0.161 \\
    & 1    & 0.497  & 0.566  & 0.580  & --1.260  & --1.130  & --1.169  & 0.173  & 0.153  & 0.176 \\
    & 2    & 0.510  & 0.532  & 0.555  & --1.520  & --1.201  & --1.601  & 0.153  & 0.175  & 0.251 \\
    & 5    & 0.487  & 0.502  & 0.637  & --1.389  & --1.368  & --0.895  & 0.163  & 0.202  & 0.279 \\
    & 10   & 0.343  & 0.584  & 0.655  & --2.370  & --1.180  & --0.895  & 0.373  & 0.228  & 0.254 \\
    & 20   & 0.358  & 0.601  & 0.515  & --2.505  & --1.190  & --0.989  & 0.537  & 0.170  & 0.245 \\
    & 50   & 0.386  & 0.646  & 0.580  & --2.224  & --1.419  & --1.450  & 0.344  & 0.182  & 0.198 \\
    \bottomrule
  \end{tabular}%
\end{table}

The $(\bone=0, \btwo=0.5)$ configuration achieves the lowest maximum
drawdown of the entire combined experiment (--2.63\%), with Calmar
0.538 and $\CVaR$ --1.107\%.  These two adjacent $\btwo$ values
correspond to the ``capital-preservation'' range for \Dirac{} at
$\bone = 0$: both configurations deliver MDD below --3.5\% while
retaining Sharpe above 0.72.  Portfolio concentration at
$(\bone=0, \btwo=0.5)$ is also relatively contained for \Dirac{}: HHI
0.196 and CR-5 0.943.

\begin{table}[htb]
  \centering
  \caption{\Dirac{} best configurations versus best classical
    configurations in the joint sweep.}
  \label{tab:dirac_corridor}
  \footnotesize \setlength{\tabcolsep}{4pt}
  \begin{tabular}{@{}lccccc@{}}
    \toprule
    Metric
    & \Dirac{} $(0,0.5)$  & \Dirac{} $(0,1)$  & \Gurobi{} $(0,1)$  & \Gurobi{} $(0,50)$  & \SAC{} $(1,0.5)$ \\
    \midrule
    Ann.\ Return (\%)      & 1.39     & 1.70     & 0.84     & 1.09     & 1.34 \\
    Ann.\ Volatility (\%)  & 1.96     & 2.23     & 1.53     & 1.57     & 2.23 \\
    Sharpe Ratio           & 0.721    & 0.760    & 0.565    & 0.715    & 0.602 \\
    Sortino Ratio          & 0.804    & 0.841    & 0.583    & 0.741    & 0.612 \\
    Realized Skewness      & --0.014  & --0.009  & +0.263   & --0.112  & --0.143 \\
    $\CVaR$ (\%)           & --1.107  & --1.278  & --0.863  & --0.911  & --1.363 \\
    Max Drawdown (\%)      & --2.63   & --3.47   & --5.37   & --4.95   & --6.99 \\
    Calmar Ratio           & 0.538    & 0.567    & 0.190    & 0.257    & 0.233 \\
    HHI                    & 0.196    & 0.340    & 0.163    & 0.187    & 0.132 \\
    CR-5                   & 0.943    & 0.895    & 0.842    & 0.792    & 0.700 \\
    \bottomrule
  \end{tabular}%
\end{table}

Exhibit~\ref{tab:dirac_corridor} provides a granular metric comparison
between \Dirac{}'s optimal configurations and the leading alternatives
from the classical pipelines within the joint hyperparameter sweep.
Relative to \Gurobi{}'s best-Sharpe configuration located at
$(\bone=0, \btwo=50)$ (Sharpe 0.715), \Dirac{} at $(\bone=0, \btwo=1)$
delivers a Sharpe premium of +0.045, a Sortino premium of +0.100, a
Calmar premium of +0.310, and an attenuation in Maximum Drawdown (MDD)
of 1.48 percentage points. These risk-adjusted gains are achieved
alongside a minor trade-off: a $\CVaR$ differential of --0.367
percentage points and a significantly higher Herfindahl-Hirschman
Index (0.340 vs.\ 0.187).

Consequently, for unconstrained return-seeking mandates where mild
tail-risk expansion can be absorbed, this performance profile
decisively favors the \Dirac{} pipeline. Conversely, for institutional
mandates where strict Conditional Value-at-Risk or top-5 concentration
ratio (CR-5) ceilings are binding, \Gurobi{} configurations such as
$(\bone=0, \btwo=1)$ or $(\bone=0, \btwo=50)$ remain the preferred
operational choice.

At the specific joint setting $(\beta_1 = 1, \beta_2 = 10)$, \Dirac{}
achieves a $\CVaR$ of --0.895\%, marking the premier tail-risk
attenuation within the $\beta_1 = 1$ sub-regime, accompanied by a
Sharpe ratio of 0.655, an MDD of --3.41\%, and a Calmar ratio of
0.381. This configuration warrants serious consideration as a
deployment candidate for risk-adjusted mandates that can tolerate
moderate portfolio concentration (HHI of 0.254, CR-5 of 0.933) while
capturing the robust tail properties characteristic of a classical
\Gurobi{}-class solver.

\Gurobi{}'s most prominent advantage across the joint sweep is its
deterministic tail-risk control. Its optimal $\CVaR$ of --0.863\% at
$(\bone=0, \btwo=1)$ is the lowest recorded across either sweep, an
outcome underpinned by a broad factor distribution where accruals,
investment, and quality share leadership at comparable weights. At
$(\bone=0, \btwo=50)$, \Gurobi{}'s peak Sharpe configuration in the
joint sweep (0.715), the $\CVaR$ remains highly contained at
--0.911\%, with an HHI of 0.187 driven compositionally by quality
(17.1\%), low-leverage (11.3\%), and seasonality (10.9\%).

Across all 16 joint configurations, \Gurobi{}’s HHI is tightly bounded
between 0.135 and 0.228, confirming a structural concentration
stability unattainable by stochastic or heuristic alternatives. For
example, under $(\beta_1 = 0, \beta_2 = 20)$, a region where the
\SAC{} pipeline undergoes severe performance degradation, \Gurobi{}
preserves a Sharpe ratio of 0.617, a $\CVaR$ of --1.230\%, and an HHI
of $0.140$. This robustness is supported in part by the exact
branch-and-bound solver, which enforces feasibility constraints and
prevents optimization divergence under extreme penalty regimes.

The region defined by $\beta_1 = 0$ and $\beta_2 \geq 10$ constitutes
a zone of structural degeneracy for the \SAC{} pipeline. The
underlying failure mechanism is rooted in objective-function
misspecification: absent a binding primary $\beta_1$ penalty anchor,
the policy gradient regularizes excessively toward low-activity
factors irrespective of their marginal return contribution, thereby
collapsing portfolio diversification entirely.

At $(\beta_1 = 0, \beta_2 = 20)$, the optimization collapses into a
concentrated two-factor allocation comprising low-leverage (30.0\%)
and investment (27.9\%), which jointly account for 57.9\% of total
portfolio weight. This structural concentration precipitates severe
performance deterioration: the Sharpe ratio drops to 0.088, MDD
expands to --14.64\%, $\CVaR$ deepens to --2.237\%, HHI rises to
0.593, and CR-5 reaches 0.985. Pushing the secondary penalty to
$(\bone=0, \btwo=50)$ drives CR-5 to 0.996, degenerating the output
into effectively a single-factor portfolio. A similar mode collapse
recurs at $(\beta_1 = 1, \beta_2 = 20)$, where the low-risk factor
alone commands 45.5\% of the allocation, the highest single-factor
weight observed across all 48 grid configurations, yielding a $\CVaR$
of --2.505\% and an HHI of 0.537.  Empirically, the parameter spaces
defined by $\beta_2 \geq 10$ under $\beta_1 = 0$, and
$\beta_2 \geq 20$ under $\beta_1 = 1$, must be strictly excluded from
any production \SAC{} operating specification to prevent catastrophic
out-of-sample drawdowns.

\subsection{Unified Efficient Frontier Comparison}

\begin{figure}[!htb]
  \centering

  \begin{subfigure}{0.48\textwidth}
    \centering
    \includegraphics[width=\textwidth]{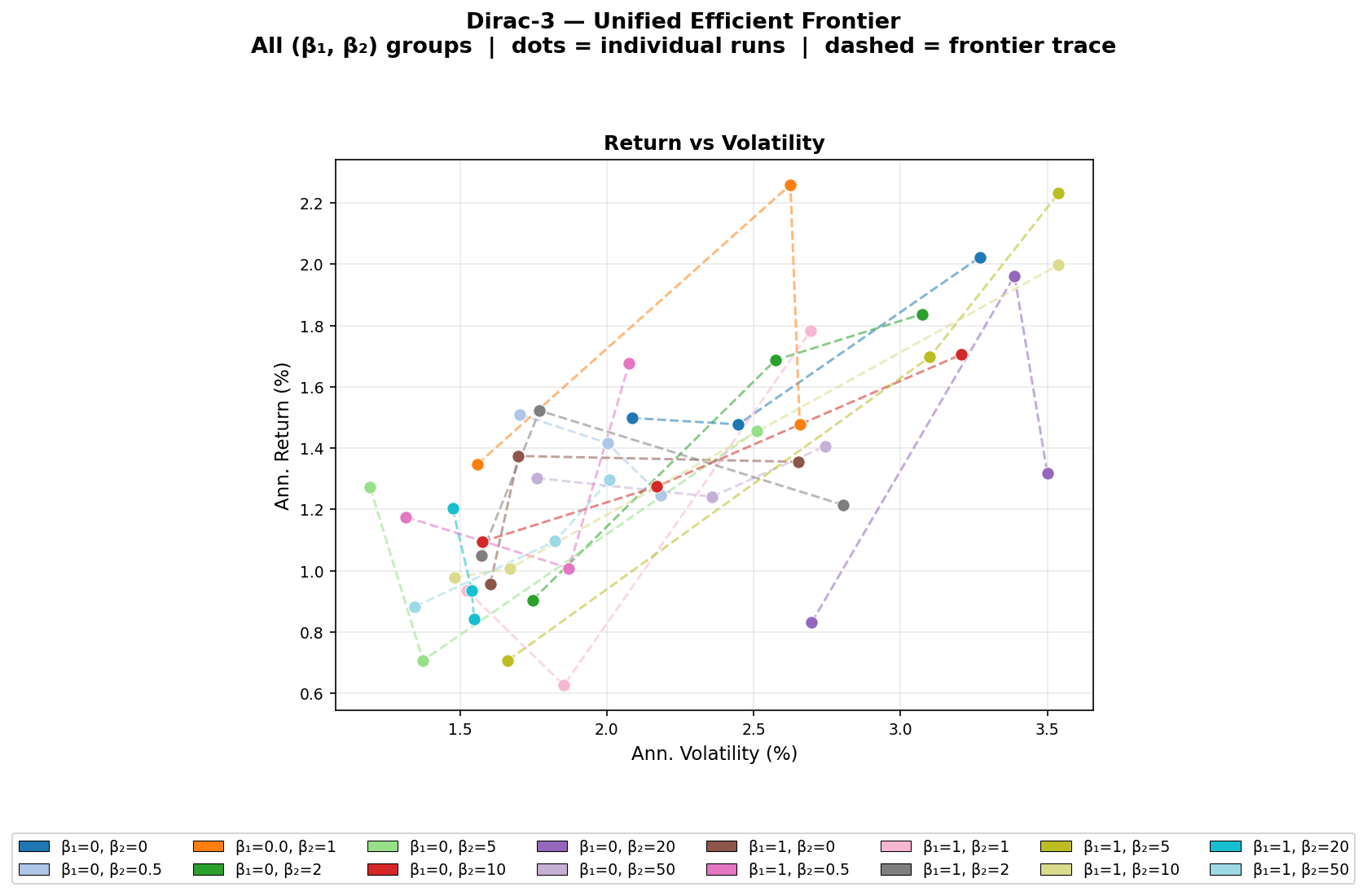}
    \caption{\Dirac{}}
    \label{fig:frontier_dirac}
  \end{subfigure}
  \hfill
  \begin{subfigure}{0.48\textwidth}
    \centering
    \includegraphics[width=\textwidth]{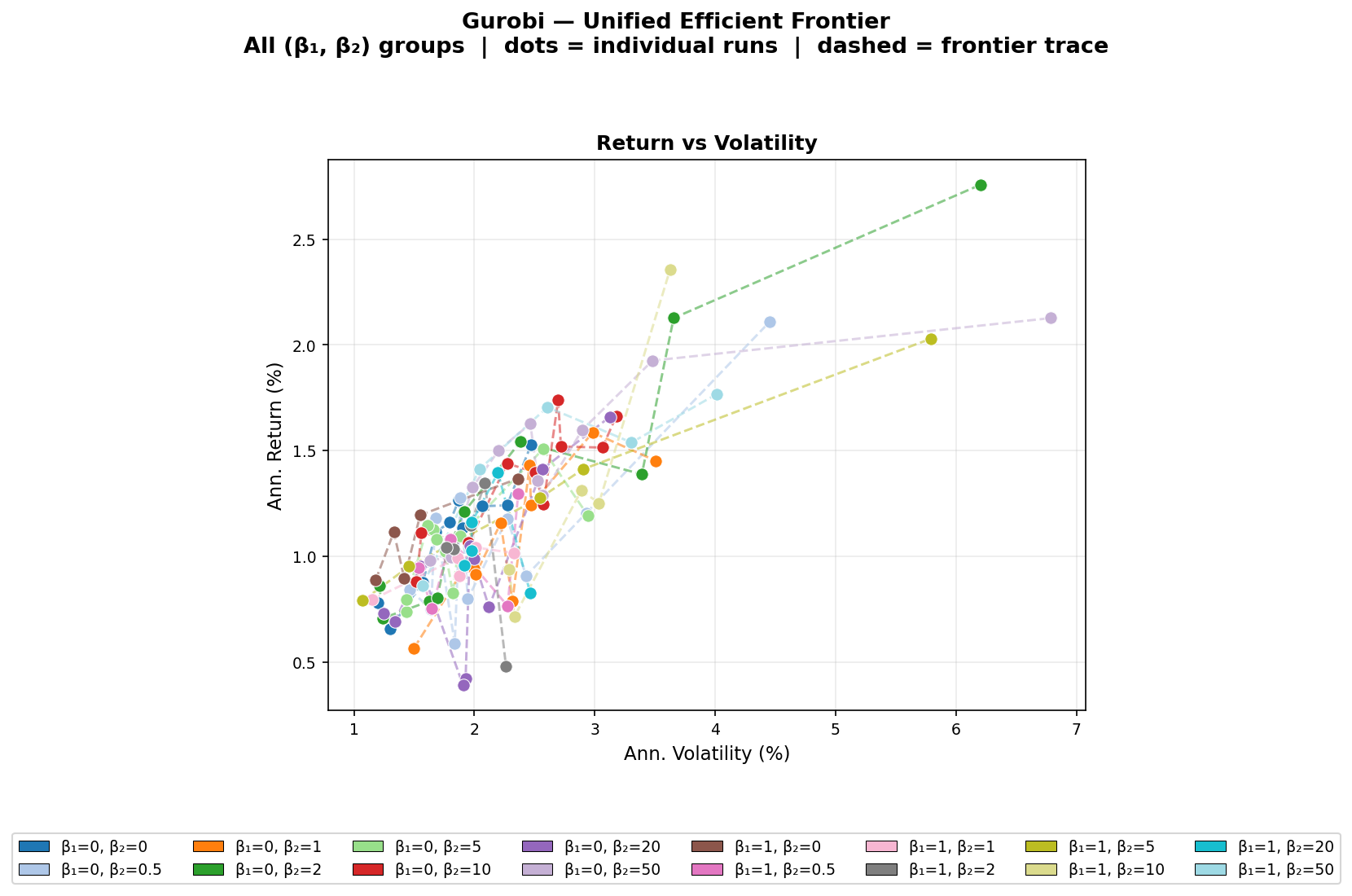}
    \caption{\Gurobi{}}
    \label{fig:frontier_gurobi}
  \end{subfigure}

  \vspace{1em}

  \begin{subfigure}{0.48\textwidth}
    \centering
    \includegraphics[width=\textwidth]{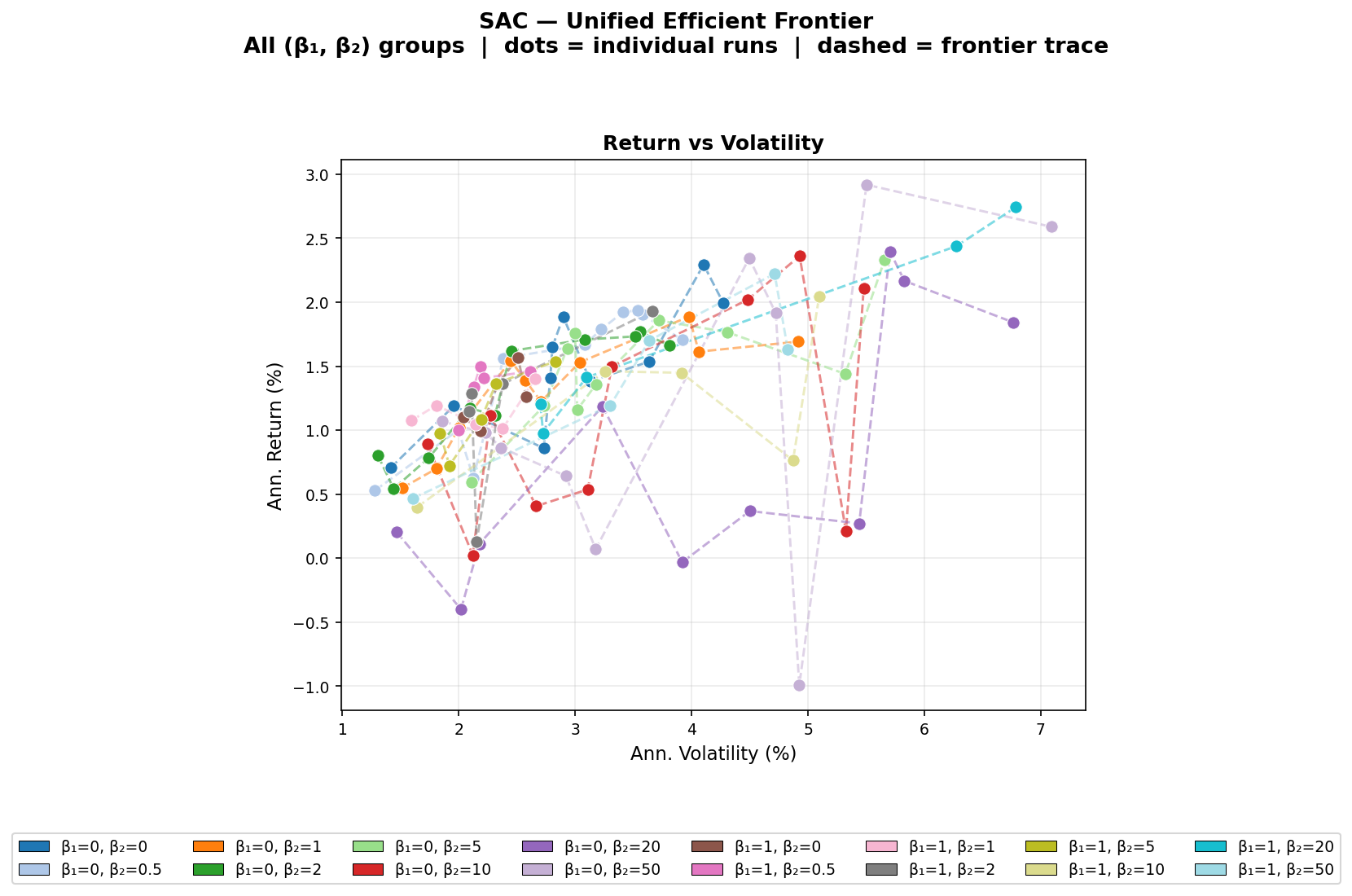}
    \caption{\SAC{}}
    \label{fig:frontier_sac}
  \end{subfigure}

  \caption{Unified efficient frontiers across all $(\bone,\btwo)$
    configurations for \Dirac{} (a), \Gurobi{} (b), and \SAC{} (c).
    Individual seed runs are shown as dots, configuration means as
    diamonds, and dashed lines trace the configuration sequence.}
  \label{fig:unified_frontiers}
\end{figure}

Exhibit~\ref{fig:unified_frontiers} compares the unified efficient
frontiers across all $(\bone, \btwo)$ configurations for the three
architectures. The probabilistic nature of \Dirac{}
(Exhibit~\ref{fig:unified_frontiers}a) leads to a wider dispersion
around the return--volatility spectrum, though it is able to explore
and locate optimal points aligning with its peak Sharpe and Calmar
ratios.  In contrast, \Gurobi{} (Exhibit~\ref{fig:unified_frontiers}b)
displays tightly clustered points, reflecting its lower cross-seed
variance and robust reproducibility. \SAC{}
(Exhibit~\ref{fig:unified_frontiers}c) demonstrates severe
instability, evidenced by wide vertical dispersion at comparable
volatility levels and negative-return outliers corresponding to the
collapse region at $\bone=0, \btwo\ge10$.

\subsection{Concentration and Portfolio Composition}

Portfolio concentration differs substantially and systematically
across the three pipelines. Exhibits~\ref{tab:conc_beta1}
and~\ref{tab:conc_joint} report HHI and CR-5 across the $\bone$ and
joint sweeps, respectively.

\begin{table}[!htb]
  \centering
  \caption{Portfolio concentration across the $\bone$ sweep (seed
    means).}
  \label{tab:conc_beta1}
  \small
  \begin{tabular}{@{}lcccccc@{}}
    \toprule
    & \multicolumn{3}{c}{HHI} & \multicolumn{3}{c}{CR-5} \\
    \cmidrule(lr){2-4} \cmidrule(lr){5-7}
    $\bone$ & \SAC{} & \Gurobi{} & \Dirac{} & \SAC{} & \Gurobi{} & \Dirac{} \\
    \midrule
    0    & 0.138 & 0.139 & 0.192 & 0.686 & 0.719 & 0.860 \\
    0.5  & 0.134 & 0.161 & 0.281 & 0.727 & 0.752 & 0.936 \\
    1    & 0.146 & 0.136 & 0.218 & 0.725 & 0.715 & 0.834 \\
    2    & 0.229 & 0.156 & 0.210 & 0.723 & 0.821 & 0.898 \\
    5    & 0.172 & 0.157 & 0.225 & 0.724 & 0.816 & 0.929 \\
    10   & 0.119 & 0.141 & 0.212 & 0.662 & 0.735 & 0.965 \\
    20   & 0.113 & 0.170 & 0.240 & 0.642 & 0.782 & 0.958 \\
    50   & 0.141 & 0.137 & 0.245 & 0.707 & 0.674 & 0.945 \\
    \midrule
    Mean & 0.150 & 0.152 & 0.228 & 0.700 & 0.752 & 0.916 \\
    \bottomrule
  \end{tabular}
\end{table}

\begin{table}[!htb]
  \centering
  \caption{HHI and CR-5 across the joint $(\bone,\btwo)$ sweep.}
  \label{tab:conc_joint}
  \small
  \begin{tabular}{@{}lcccccc@{}}
    \toprule
    & \multicolumn{3}{c}{HHI} & \multicolumn{3}{c}{CR-5} \\
    \cmidrule(lr){2-4} \cmidrule(lr){5-7}
    Config & \SAC{} & \Gurobi{} & \Dirac{} & \SAC{} & \Gurobi{} & \Dirac{} \\
    \midrule
    $(0,0)$    & 0.201 & 0.148 & 0.353 & 0.758 & 0.758 & 0.957 \\
    $(0,0.5)$  & 0.154 & 0.135 & 0.196 & 0.753 & 0.719 & 0.943 \\
    $(0,1)$    & 0.177 & 0.163 & 0.340 & 0.721 & 0.842 & 0.895 \\
    $(0,2)$    & 0.154 & 0.157 & 0.268 & 0.727 & 0.813 & 0.883 \\
    $(0,5)$    & 0.200 & 0.140 & 0.235 & 0.811 & 0.736 & 0.966 \\
    $(0,10)$   & 0.333 & 0.170 & 0.245 & 0.948 & 0.804 & 0.965 \\
    $(0,20)$   & 0.593 & 0.140 & 0.281 & 0.985 & 0.723 & 0.919 \\
    $(0,50)$   & 0.520 & 0.187 & 0.225 & 0.996 & 0.792 & 0.955 \\
    \midrule
    $(1,0)$    & 0.153 & 0.138 & 0.181 & 0.735 & 0.706 & 0.825 \\
    $(1,0.5)$  & 0.132 & 0.163 & 0.161 & 0.700 & 0.851 & 0.830 \\
    $(1,1)$    & 0.173 & 0.153 & 0.176 & 0.707 & 0.784 & 0.880 \\
    $(1,2)$    & 0.153 & 0.175 & 0.251 & 0.734 & 0.771 & 0.966 \\
    $(1,5)$    & 0.163 & 0.202 & 0.279 & 0.792 & 0.885 & 0.966 \\
    $(1,10)$   & 0.373 & 0.228 & 0.254 & 0.935 & 0.955 & 0.933 \\
    $(1,20)$   & 0.537 & 0.170 & 0.245 & 0.976 & 0.875 & 0.933 \\
    $(1,50)$   & 0.344 & 0.182 & 0.198 & 0.979 & 0.869 & 0.923 \\
    \bottomrule
  \end{tabular}
\end{table}

Cross-sectional comparison reveals three distinct characteristics
across the optimization engines. First, \Dirac{} portfolios are
consistently more concentrated than the classical baselines,
exhibiting a mean HHI of 0.228 in the $\beta_1$ sweep, compared to
0.150 for \SAC{} and $0.152$ for \Gurobi{}. Furthermore, \Dirac{}'s
top-5 concentration ratio (CR-5) averages 0.916, indicating that the
five largest factor allocations routinely capture over 90\% of total
portfolio weight. Institutional practitioners operating under strict
mandate-mandated CR-5 ceilings should note that no \Dirac{}
configuration across either sweep achieves a CR-5 below 0.825.
Second, \Gurobi{} maintains the most stable diversification profile:
its HHI is tightly bounded between 0.135 and 0.228 across the entire
joint sweep, with no single factor exceeding 17.7\% allocation in any
configuration. This consistency is directly attributable to the exact
branch-and-bound solver's ability to satisfy penalty objectives while
preserving portfolio breadth.  Third, \SAC{} exhibits a bimodal
concentration structure: it remains well-controlled (HHI between 0.113
and 0.229) in low-$\beta_2$ regimes, but experiences severe
concentration (yielding an HHI of up to 0.593) when $\beta_2 \ge 10$
under $\beta_1 = 0$.

Across both sweeps, the three pipelines exhibit distinct factor
allocation patterns that remain largely invariant within each solver
architecture.  High-performing \Dirac{} configurations consistently
converge on a concentrated handful of factors. At $\beta_1 = 1$ (the
optimal Calmar setting in the primary sweep), the three largest
allocations are accruals (20.5\%), investment (14.5\%), and
low-leverage (10.2\%). At the global optimum $(\beta_1=0, \beta_2=1)$,
accruals (25.9\%) and investment (19.6\%) jointly command 45.5\% of
total weight. At $(\bone=0, \btwo=0.5)$, the configuration achieving
the lowest MDD, investment and low-leverage each capture 20.0\%. Note
that these specific factors, like accruals and investment, are
well-documented high-return, low-tail-risk anomalies within the JKP
data library. The photonic solver's stochastic search over the
quadratic unconstrained binary optimization (QUBO) landscape appears
to discover and exploit them more reliably than deterministic
alternatives.

Deviations from this compositional pattern serve as reliable basis for
operational diagnostics. When setting $\beta_1 = 0.5$, momentum spikes
to 20.6\% of total weight, coinciding directly with the observed
volatility expansion and turnover anomalies. Similarly, at
$(\bone=0, \btwo=20)$, the accruals allocation becomes fragmented and
isolated, degrading Sharpe ratio to 0.421 and MDD to
--11.09\%. Consequently, tracking \Dirac{} factor composition provides
a valuable real-time operational signal: an undue dominance of
momentum or low-risk factors indicates that the portfolio has drifted
into a degraded transition-zone regime.

Factor weights evolve smoothly across penalty configurations,
completely avoiding the single- or two-factor dominance observed in
optimal \Dirac{} or failure-state \SAC{} portfolios. At
$(\bone=0, \btwo=1)$ (optimal $\CVaR$), accruals, investment, and
quality share leadership at comparable weights. At
$(\bone=0, \btwo=50)$ (peak Sharpe in the joint sweep), quality
(17.1\%), low-leverage (11.3\%), and seasonality (10.9\% ) co-lead the
allocation. This compositional breadth directly underpins \Gurobi{}'s
narrow $\CVaR$ bandwidth and minimal cross-seed variance across
specifications.

In the $\beta_1 = 0$ regime, portfolio composition degrades
progressively as the secondary penalty $\beta_2$ increases. At
$\beta_2 = 0$, the portfolio is broadly diversified across 13 factors,
with no single allocation exceeding 14.8\%. However, at
$\beta_2 = 20$, two factors, low-leverage (30.0\% ) and investment
(27.9\%), absorb 57.9\% of total weight. A similarly severe mode
collapse occurs under $\beta_1 = 1$; at $(1, 20)$, the low-risk factor
alone commands 45.5\% of the allocation, representing the highest
single-factor concentration recorded across all 48 grid
configurations. Lacking the explicit constraint-handling mechanisms of
the QUBO or MILP frameworks, the reinforcement learning policy
gradient regularizes excessively toward low-activity factors under
heavy penalty pressures, regardless of their underlying marginal
return contributions.

Exhibit~\ref{tab:conc_summary} synthesizes concentration profiles by
pipeline.  The key observation is that \Dirac{}'s persistently high
CR-5 (typically 0.83--0.97) may conflict with explicit diversification
mandates.  Practitioners with binding CR-5 limits should use \Gurobi{}
as the primary pipeline, or apply post-optimization weight capping to
\Dirac{} allocations.

\begin{table}[htbp]
  \centering
  \caption{Concentration profile summary by pipeline across both
    sweeps.}
  \label{tab:conc_summary}
  \small
  \begin{tabular}{@{}lcccl@{}}
    \toprule
    & HHI Range     & HHI Mean  & CR-5 Range    & Dominant failure mode \\
    \midrule
    \SAC{}     & 0.113--0.593  & 0.211     & 0.642--0.996  & High-$\btwo$ collapse at $\bone = 0$ \\
    \Gurobi{}  & 0.135--0.228  & 0.163     & 0.674--0.955  & None observed \\
    \Dirac{}   & 0.161--0.353  & 0.234     & 0.825--0.966  & Momentum/low-risk tilt \\
    \bottomrule
  \end{tabular}
\end{table}

\section{Discussions}
\label{sec:discussion}

\subsection{Performance Characteristics by Pipeline}

Across 48 configurations and 164 months, the three pipelines exhibit
distinct and largely non-overlapping comparative advantages and
weaknesses.  \Dirac{} achieves the highest values for Sharpe ratio,
Calmar ratio, and maximum drawdown protection at its respective
optima, but these results are concentrated in two narrow penalty
windows: $\bone \in \{1, 2\}$ in the primary sweep, and
$\btwo \in \{0.5, 1\}$ at $\bone = 0$ in the joint sweep. The
underlying mechanism, as revealed by the composition analysis, is that
the photonic solver preferentially selects portfolios concentrated in
accruals and investment (factors with documented positive-return
persistence) when the penalty regime is correctly calibrated. Outside
this calibration window, performance degrades materially.  The
concentration-performance link is a defining characteristic of
\Dirac{}: its advantages are inseparable from its concentration
tendency, and both are configuration-sensitive.

\Gurobi{}'s advantage is in tail-risk control and reproducibility. It
achieves the best $\CVaR$ in the combined sweep (--0.863\%), and its
cross-seed variance is the narrowest of the three architectures. These
properties are mechanistically linked to the deterministic feasibility
certificate of the branch-and-bound solver, which prevents
compositional collapse at extreme penalty settings.  \Gurobi{}'s
compositional stability under all 16 joint configurations (HHI never
exceeding 0.228 even at $(\bone = 0, \btwo = 20)$ where \SAC{}
collapses to HHI 0.593) is unique among the three pipelines. For
portfolio managers who prioritize predictable and stable performance,
\Gurobi{} is the most suitable primary pipeline even if its peak
Sharpe and Calmar lag \Dirac{} at optimal settings.

\SAC{}'s utility is limited to low-penalty regimes. As a
return-seeking pipeline with moderate $\btwo$, it is competitive on
raw return figures but consistently inferior on tail-risk and
seed-stability metrics. Its structural disadvantage relative to the
\QUBO{}-augmented pipelines stems from the absence of the factor
selection layer: the policy gradient is regularized under large
penalties toward low-activity factors (low-risk, low-leverage) not
because these factors offer superior risk-adjusted return but because
they generate the smallest step-change in the penalty objective. The
result is a portfolio that concentrates for the wrong reasons and
generates poor risk-adjusted outcomes.

\subsection{Mandate-Specific Recommendations}

Exhibit~\ref{tab:recommendations} consolidates solver-configuration
recommendations by mandate type.  These are derived directly from the
empirical evidence in preceding analyses and should be treated as
starting points for further validation in out-of-sample testing.

\begin{table}[!htb]
  \centering
  \caption{Recommended solver-configuration pairs by mandate type.}
  \label{tab:recommendations}
  \small
  \begin{tabular}{@{}llp{5.5cm}@{}}
    \toprule
    Mandate                         & Configuration                            & Key metrics \\
    \midrule
    Return-seeking, balanced        & \Dirac{} $(0,1)$                         & Sharpe 0.760, Calmar 0.567, MDD --3.47\%. Best overall risk-adjusted configuration. \\[2pt]
    Lowest drawdown                 & \Dirac{} $(0,0.5)$                       & MDD --2.63\% (best in both sweeps), Calmar 0.538. \\[2pt]
    Best tail-risk control          & \Gurobi{} $(0,1)$                        & $\CVaR$ --0.863\% (best in both sweeps), seed-stable. \\[2pt]
    Balanced risk, classical        & \Gurobi{} $\bone\!=\!5$                  & Sharpe 0.718, $\CVaR$ --1.049\%, MDD --4.33\%. Best primary-sweep Gurobi result. \\[2pt]
    Maximum reproducibility         & \Gurobi{} $\bone\!=\!20$                 & Return std 2.10pp; tightest cross-seed variance in either sweep. \\[2pt]
    CVaR at moderate concentration  & \Dirac{} $(1,10)$                        & $\CVaR$ --0.895\%, Calmar 0.381; warrants further out-of-sample evaluation. \\[2pt]
    Avoid (concentration risk)      & \Dirac{} $\bone\!=\!0.5$                 & Transition instability; momentum tilt, HHI 0.281, turnover 0.044. Exclude from all operating sets. \\[2pt]
    Avoid (concentration risk)      & \Dirac{} $(0,0)$                         & HHI $= 0.353$, CR-5 = 0.957; extreme concentration without commensurate return. \\[2pt]
    Avoid (catastrophic failure)    & \SAC{} $\bone\!=\!0$, $\btwo\!\geq\!10$  & Sharpe $<$ 0.31, HHI $>$ 0.33, MDD up to --14.64\%. Non-deployable. \\[2pt]
    Avoid (high seed risk)          & \SAC{} $\bone\!\geq\!10$                 & Sharpe std 0.206 at $\bone=20$. Cross-seed range 0.55 Sharpe points. \\
    \bottomrule
  \end{tabular}
\end{table}

\subsection{Operational Monitoring \& Limitations}

Given the configuration sensitivity of the \Dirac{} pipeline and the
structural failure modes of \SAC{}, live production deployment
necessitates explicit risk-monitoring protocols:
\begin{enumerate}
\item Composition alerts for \Dirac{}: Any operational portfolio in
  which momentum or low-risk emerges as the single largest factor
  weight should trigger an immediate regime review. This compositional
  signature has consistently preceded performance degradation across
  both hyperparameter sweeps, providing a leading indicator of regime
  instability before quantitative drawdown metrics deteriorate.
\item HHI thresholding for \Dirac{}: Configurations where the
  Herfindahl-Hirschman Index exceeds 0.30 (such as at
  $(\bone=0, \btwo=0)$ and $(\bone=0, \btwo=1)$) require
  post-optimization weight capping or tactical rebalancing toward an
  anchor benchmark. Imposing a composite concentration ceiling of HHI
  $<$ 0.25 effectively isolates the vast majority of \Dirac{}'s
  well-performing specifications while screening out
  mandate-non-compliant tail concentrations.
\item Single-factor exposure limits for \SAC{}: For
  reinforcement-learning-driven pipelines, any single factor breaching
  30\% of total portfolio weight should serve as a hard stop. In both
  identified failure configurations (low-leverage capturing 30.0\% at
  $(\bone=0, \btwo=20)$ and low-risk reaching 45.5\% at
  $(\bone=1, \btwo=20)$) this factor concentration threshold was
  breached immediately prior to catastrophic metric collapse.
\end{enumerate}

Finally, three primary considerations of this study merit explicit
acknowledgment.  The first concerns historical path dependency. The
164-month backtest window represents a single realization of
historical market paths. Assessing the out-of-sample stability of
solver performance rankings across distinct macro-regimes (e.g.,
structural interest rate shifts or prolonged liquidity crunches)
remains an essential area for future research.  The second
consideration concerns hardware evolution. The \Dirac{} evaluations
reflect a specific generational iteration of photonic
hardware. Algorithmic efficiency and sample convergence may scale
materially on future device generations offering expanded entropy
bandwidth and reduced sampling noise. The final consideration is about
execution friction and latency. The economic impact of cloud API
latency and execution friction associated with the photonic solver has
not been modeled. For higher-frequency mandates or
capacity-constrained strategies, network round-trip times could
introduce implementation drag that alters the net-of-cost performance
frontier.

\section{Conclusion}
\label{sec:conclusion}

This study provides a rigorous, large-scale empirical evaluation of
three distinct optimization paradigms, photonic quantum annealing
(\Dirac{}), exact mixed-integer programming (\Gurobi{}), and
stochastic reinforcement learning (\SAC{}), applied to constrained
equity factor allocation. By mapping performance across 48
hyperparameter configurations over a 164-month horizon, our findings
demonstrate that portfolio optimization topology is profoundly
architecture-dependent. Rather than identifying a single universally
dominant solver, the empirical evidence reveals a clear institutional
trade-off between heuristic-driven alpha capture, deterministic
tail-risk resilience, and stochastic policy regularization.

At the frontier of risk-adjusted performance, \Dirac{} establishes
superior metrics, delivering a peak Sharpe ratio of $0.760$, a Calmar
ratio of $0.567$, and a maximum drawdown attenuation of
$-2.63\%$. However, these efficiencies are strictly bounded within
narrow operational sweet spots and are compromised by high factor
concentration (mean HHI $>0.228$; CR-5 $>0.90$), driven by the
photonic solver's selection of persistent, quality-adjacent signals
like accruals and investment. In contrast, \Gurobi{} serves as the
benchmark for institutional stability. By guaranteeing feasibility via
branch-and-bound solving, \Gurobi{} achieves superior conditional
Value-at-Risk ($\CVaR$ of $-0.863\%$), minimal cross-seed variance,
and a tightly bounded compositional breadth across the entire grid,
degrading gradually under penalty stress rather than experiencing
sudden collapse. Meanwhile, \SAC{} captures high nominal returns in
unconstrained regimes but exhibits severe structural fragility. In the
absence of a binding primary volatility anchor, high secondary
skewness penalties trigger objective-function misspecification,
forcing the policy gradient to regularize excessively toward
low-activity factors and resulting in catastrophic concentration
failure.

For institutional practitioners and quantitative researchers
evaluating quantum and advanced machine learning solvers for live
deployment, these findings translate into concrete operational
guidelines. Portfolio managers must carefully match solver
architecture to the constraints of their investment mandate, choosing
\Dirac{} for return-seeking mandates where maximizing Sharpe and
Calmar ratios justifies accepting higher portfolio concentration and
active post-optimization weight capping. Conversely, fiduciary
mandates bound by strict drawdown limits, regulatory diversification
ceilings, and requirements for reproducible, cross-seed stability are
best served by \Gurobi{}, whereas \SAC{} deployments must be strictly
restricted to tightly regularized regimes unless backed by robust
ensemble averaging and explicit constraint enforcement. Furthermore,
portfolio managers must treat hyperparameter calibration as a primary
risk factor, recognizing that optimization engines relying on
heuristic sampling or policy gradients require dedicated validation
layers to prevent transitions into degraded parameter zones.

To safeguard live execution, quantitative systems should implement
real-time compositional circuit breakers. Because factor-weight
concentration consistently precedes quantitative metric degradation,
live systems should monitor portfolio composition continuously,
triggering automated reviews whenever a \Dirac{} configuration is
dominated by momentum or low-risk factors, or whenever a \SAC{}
allocation breaches a $30\%$ single-factor threshold. Finally, while
\Dirac{} demonstrates clear statistical advantages in out-of-sample
risk-adjusted returns, live implementation must weigh these against
cloud API latency, rebalancing frequency, and turnover constraints.

Ultimately, this study demonstrates that photonic annealing offers a
viable, high-performing alternative to classical solvers in
constrained multi-factor equity portfolio construction, provided its
configuration sensitivity is actively managed. As quantum hardware
scales in entropy bandwidth and sampling fidelity, the integration of
QUBO-based solvers into institutional asset management pipelines will
increasingly depend on rigorous operational guardrails, bridging the
gap between mathematical optimization and robust real-world execution.

\clearpage
\singlespacing \printbibliography

\end{document}